\documentclass{IEEEtran}
\usepackage{cite}
\usepackage{amsmath,amssymb,amsfonts}
\usepackage{algorithm}
\usepackage{algorithmicx}
\usepackage{graphicx}
\usepackage{textcomp}
\usepackage{xcolor}
\usepackage[font=footnotesize]{subfig}
\graphicspath{{figures/}}
\def\BibTeX{{\rm B\kern-.05em{\sc i\kern-.025em b}\kern-.08em
    T\kern-.1667em\lower.7ex\hbox{E}\kern-.125emX}}

\begin{document}
\title{Generative Artificial Intelligence Assisted Multi-modal Semantic Extraction for Noma-based Image Transmissions}
\author{Songhan Zhao, Shimin Gong, Bo Gu, Hongyang Du, Xidong Mu, Zehui Xiong, and Yuming Fang
\thanks{
Songhan Zhao, Shimin Gong, and Bo Gu are with the School of Intelligent Systems Engineering, Sun Yat-sen University, China (e-mails: zhaosh55@mail2.sysu.edu.cn, \{gongshm5, gubo\}@mail.sysu.edu.cn). 

Hongyang Du is with the Department of Electrical and Electronic Engineering, University of Hong Kong, Hong Kong, China (e-mail: duhy@eee.hku.hk). 

Xidong Mu and Zehui Xiong are with the School of Electronics, Electrical Engineering and Computer Science (EEECS), Queen's University Belfast, U.K. (e-mail:\{x.mu, z.xiong\}@qub.ac.uk). 

Yuming Fang is with the School of Information Management, Jiangxi University of Finance and Economics, China (e-mail:leo.fangyuming@foxmail.com).
}
}
\maketitle
\thispagestyle{empty}

\begin{abstract}
In this paper, we investigate a generative artificial intelligence (GAI)-assisted semantic communication framework for non-orthogonal multiple access (NOMA)-based image transmissions. Semantic users (SUs) extract cross-modal semantic features from the raw images, which are then used for image recovery by leveraging a GAI model. The GAI enhances the generalization and recovery of semantic image transmissions, while NOMA efficiently allocates transmission capacities to SUs based on their traffic demands. Thus, the semantic extraction and transmission control jointly affect both semantic recovery performance and transmission overhead. We maximize a weighted performance of transmission latency and semantic recovery accuracy by jointly optimizing the semantic feature selection at the semantic level, as well as the receive beamforming and NOMA decoding order at the transmission level. To reduce potential redundancy in semantic features and improve optimization efficiency, we develop an importance-aware and model-driven proximal policy optimization (IM-PPO) framework. Specifically, we quantify and retain high-importance semantic features to enhance the learning efficiency of PPO, while model-based optimization methods are used to adapt the transmission control variables.  Numerical results validate that the joint adjustment of the semantic feature selection and the transmission control significantly improves the semantic recovery accuracy and the transmission latency performance. Moreover, the IM-PPO framework effectively leverages the model information to improve the learning efficiency compared to benchmark methods.
\end{abstract}
\section{Introduction}
The emergence of large-scale Internet of Things (IoT) has dramatically increased network traffic demands, posing significant challenges to existing communication infrastructures, such as spectrum scarcity and increased transmission latency~\cite{Yang-2023icst}. Many IoT applications rely heavily on image transmissions, which further puts extra pressure on wireless transmissions due to the high data traffic and stringent quality requirements of image data. Semantic communication has emerged as a promising solution to address these bottlenecks. In many scenarios, such as surveillance and remote sensing, transmitting the raw image is unnecessary because only the essential meaning needs to be conveyed. As such, unlike conventional bit-level transmissions, semantic communication extracts and transmits the key semantic information~\cite{Xu-2023jstsp}, which allows the system to reshape users’ traffic demands and efficiently adapt to varying network conditions~\cite{Luo-2022iwc}. 

Existing research has explored different techniques for enabling semantic communication. The authors in~\cite{Yang-2023jasc} employed knowledge graphs to characterize the inherent relationships among semantic entities. By filtering less relevant entities, semantic communications can focus on more essential information and achieve a lower data volume. The authors in~\cite{Xie-2021tsp} proposed a deep learning-based semantic communication (DeepSC) system that leverages joint source-channel coding to achieve efficient end-to-end semantic transmission.
Motivated by the advantages of semantic communication, its incorporation into existing wireless networks is expected to overcome their inherent limitations. For example, the authors in~\cite{Long-2025twc} investigated an unmanned aerial vehicle (UAV)-assisted semantic communication network. The UAV can serve ground users more promptly by collecting lightweight semantic information and thus significantly reduce the system age-of-information (AoI). The authors in~\cite{Mu-2023jsac} integrated semantic communication into the non-orthogonal multiple access (NOMA) system. The co-channel interference among users is alleviated with the aid of semantic communications, thus achieving higher transmission efficiency. Existing semantic communication techniques generally rely on static knowledge bases or deep learning models, which are used to extract semantic information at the transmitter and correspondingly recover it at the receiver. However, due to their task-specific design properties, they are difficult to adapt to dynamic network variations. Although the model fine-tuning can offer adaptability, it incurs considerable time and computational overhead, limiting the practicality in real-time and resource-constrained scenarios.

Generative artificial intelligence (GAI) models, such as stable diffusion model (SDM)~\cite{Zhang-2023iccv} and ChatGPT~\cite{Wu-2023IJAS}, have demonstrated strong generalization capabilities across diverse wireless scenarios by leveraging massive model parameters and extensive pre-training on large-scale datasets~\cite{Khoramnejad-2025icst}. By providing customized prompts, the GAI can generate the desired content (e.g., images) without requiring repeated retraining. Recent studies have shown that GAI can serve various roles in wireless networks, such as optimizers~\cite{guo-2024arxiv} and semantic decoders~\cite{Yang-2025twc}. The authors in~\cite{guo-2024arxiv} demonstrated that GAI enables optimization without requiring gradient information, making it suitable for problems with unknown or non-differentiable loss functions. The authors in~\cite{Yang-2025twc} employed GAI as semantic decoders by reconstructing road scenes and predicting future states based on high-level descriptions of road node positions.
Motivated by the generalization capability of GAI, we aim to incorporate GAI into semantic communications to better adapt to dynamic network conditions and diverse task requirements. However, semantic feature extraction dynamically reshapes users' traffic demands, posing challenges to conventional transmission control schemes that are typically tailored to fixed traffic patterns.

Semantic extraction aims to select high informative semantic features to improve semantic recovery of raw images, while a tailored transmission strategy should be designed to match the traffic demands of the selected semantic features. This leads to a joint optimization of semantic control and transmission control variables~\cite{Wang-2024tcom}. However, as semantic communication models often operate as black boxes, it is challenging to directly apply conventional model-based optimization methods. Deep reinforcement learning (DRL) has been regarded as an efficient solution for complex decision-making tasks under incomplete model information~\cite{Liu-2023iwc}. By interacting with the environment, DRL agents can continuously improve their decision-making abilities through accumulated experience. Thus, DRL can be an efficient method to address optimization problems in semantic communications. The authors in~\cite{Hu-2025tcom} investigated a UAV-assisted semantic communication network, where  DRL is employed to jointly optimize the size of transmitted semantic symbols and the UAV's trajectory planning. Although DRL can jointly optimize all control variables in semantic communications, it may overlook available model information, such as channel conditions and interference characteristics. Moreover, as the dimension of control variables increases, the DRL's learning process becomes more computationally intensive, which can hinder the convergence to optimal solutions.

To address the above difficulties,  we explore a GAI-assisted semantic communication framework for wireless image transmissions. Multiple semantic users (SUs) hold image content to be transmitted to the access point (AP) via the NOMA-based semantic communication. To achieve reliable semantic transmission, we extract cross-modal features from the raw image as semantic information. Textual features are extracted to enhance high-level semantic understanding, while visual features are used to preserve low-level details, jointly improving image recovery performance at the receiver. Note that different extracted semantic features impose varying semantic traffic demands among SUs. We employ NOMA to address this by adaptively assigning a decoding order that matches each SU's traffic demands. For example, SUs with heavier semantic traffic demand can be assigned a favorable decoding order to better satisfy their transmission requirements, while those with lighter semantic traffic demand can be assigned a lower decoding priority with reduced transmission overhead.

Therefore, we propose a joint transmission and semantic control (JTSC) scheme to consider both semantic recovery accuracy and transmission latency. This weighted performance is maximized by jointly optimizing the semantic feature selection strategy, the NOMA decoding order, and the AP's receive beamforming. We propose an importance-aware and model-driven proximal policy optimization (IM-PPO) framework. Based on the availability of explicit model information, the control variables are decomposed into two parts. Model-based optimization methods are applied to the transmission control variables, while model-free PPO is employed to optimize the semantic feature selection strategy. To reduce potential redundancy in cross-modal features, we design a novel cross-modal matching approach to quantify the importance of semantic features, thus pruning the search space and accelerating the PPO's learning process. The main contributions of this paper are summarized as follows:
\begin{itemize}
\item \emph{GAI-assisted Semantic Image Transmission}:
A JTSC scheme is proposed to improve both semantic fidelity and transmission efficiency. SUs adopt task-oriented semantic selection strategies, while NOMA allocates appropriate transmission capacities to meet their diverse traffic demands. We maximize the weighted performance of transmission latency and semantic recovery accuracy by jointly optimizing the semantic feature selection strategy, the NOMA decoding order, and the receive beamforming.
\item \emph{Cross-modal Matching for Semantic Feature Pruning}:
While cross-modal features enable a more fine-grained representation of the raw image, they can also potentially introduce redundant information. To address this, we design a cross-modal matching method to quantify the importance of all extracted features.  By filtering out low-importance features, we effectively reduce the feature space and thus improve the efficiency of the subsequent PPO's learning process.
\item \emph{Importance-aware and Model-driven PPO Framework}:
We propose an IM-PPO framework, where the coupled control variables are decomposed into two parts based on the availability of model information. The model-based optimization methods adapt the transmission control variables, while the model-free PPO selects semantic features from the pruned search space. Simulation results validate that the JTSC scheme effectively enhances both semantic recovery and transmission latency performance, while the IM-PPO framework significantly improves learning efficiency compared to benchmark methods.
\end{itemize}

The rest of this paper is organized as follows. Section~\ref{system-problem} presents the system model and problem formulation. Section~\ref{opti-transmit} proposes a model-based optimization method for transmission control, while Section~\ref{impor-ppo} develops an importance-aware PPO for semantic feature selection. The effectiveness of the JTSC scheme and IM-PPO framework is validated in Section~\ref{results}. Finally, Section~\ref{conclusions} concludes the paper.
\section{System Model and Problem Formulation}\label{system-problem}
\begin{figure}[t]
	\centering
	\includegraphics[width = 0.45\textwidth]{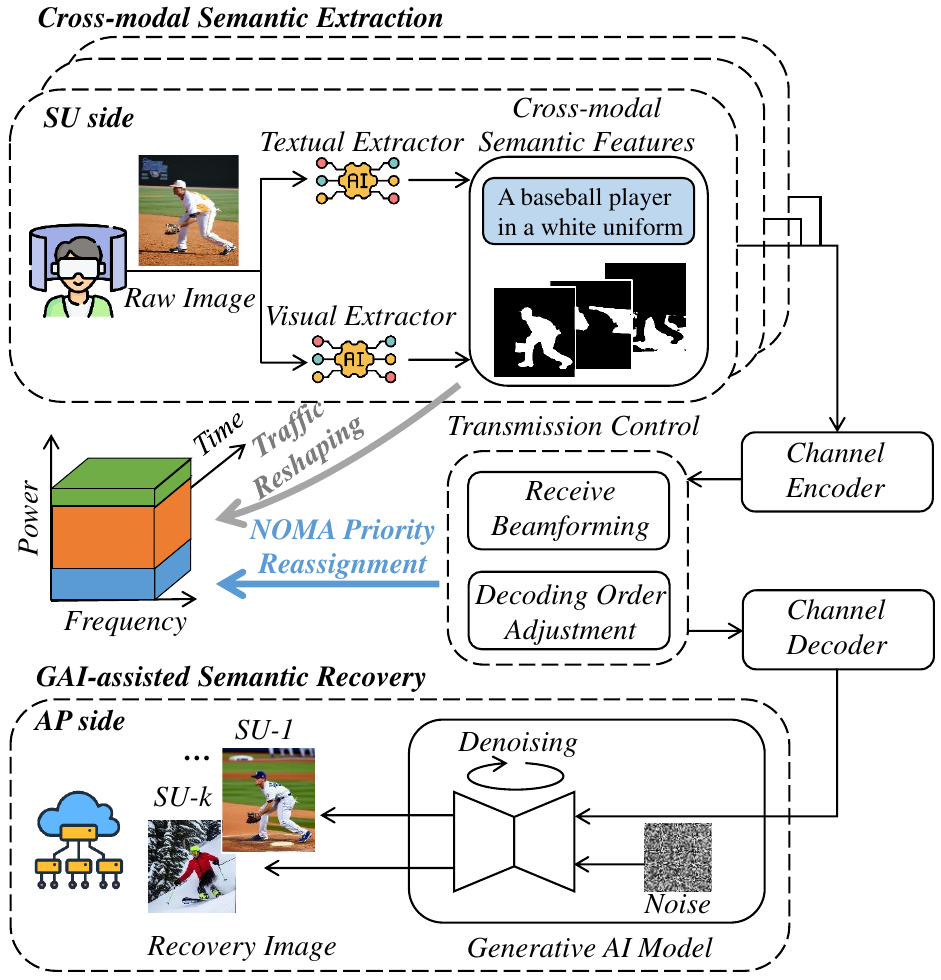}
	\caption{ GAI-assisted semantic image transmissions}\label{system-model}
\end{figure}

In Fig.~\ref{system-model}, we consider a GAI-assisted semantic image transmission, where $K$ SUs hold image data that needs to be transmitted to the AP. We consider a scenario where the SUs do not require the exact raw images but only need to convey their underlying meaning, such as in traffic monitoring. To improve transmission efficiency, SUs extract semantic information from raw images and transmit it using NOMA. The AP employs a pre-trained GAI model to recover the image data from the received semantic information. While richer semantic information can improve recovery accuracy, it also incurs increased transmission overhead due to larger data volumes. Meanwhile, SUs have different sizes of semantic features, requiring appropriate NOMA decoding orders to meet their respective traffic demands. As such, the semantic control and transmission control are inherently coupled and thus should be jointly adjusted to enhance overall system performance.
\subsection{Cross-Modal Semantic Feature Extraction}
To enable reliable semantic image transmissions, we employ both textual and visual features as cross-modal features. We employ image captions as textual features to provide human-interpretable intentions, which convey highly compressed meanings and reflect high-level objectives~\cite{Dong-2022acs}. In contrast, semantic segmentation maps are leveraged as visual features~\cite{Long-2025CVPR}, capturing fine-grained structural semantics that are difficult to solely describe by textual features.
The Unified Perceptual Parsing Network (UperNet)~\cite{Wang-2023ICETCI} and Bootstrapping Language-Image Pretraining (BLIP)~\cite{pmlr-v162-li22n} models are adopted for visual and textual feature extraction, respectively.
Let $\mathcal{L}_v$ and $\mathcal{L}_t$ represent the sets of extracted visual and textual features, respectively. The visual and textual features of SU-$k$ are represented as $\mathbf{f}_{v,k}=\{{f}_{v,k,n}\}_{n\in\mathcal{L}_v}$ and $\mathbf{f}_{t,k}=\{{f}_{t,k,n}\}_{m\in\mathcal{L}_t}$, where $f_{v,k,n}$ and $f_{t,k,m}$ denote the $n$-th visual feature and $m$-th textual feature, respectively. Given the raw image, the cross-modal features are obtained as follows:
\begin{equation}\label{semantic-extraction}
\mathbf{f}_{v,k} = \mathcal{U}(\mathbf{s}_k) \text{ and } \mathbf{f}_{t,k} = \mathcal{B}(\mathbf{s}_k),
\end{equation}
where $\mathcal{U}(\cdot)$ and $\mathcal{B}(\cdot)$ represent the mapping functions from the raw image to the visual and textual features, respectively.

The initially extracted semantic information contains comprehensive semantic features. However, some of these are redundant, resulting in additional transmission latency. To address this, we aim to select and transmit only the most informative semantic features to improve NOMA transmission efficiency while preserving semantic fidelity. Let binary variable $\boldsymbol{\beta}_k = \{ \beta_{k,i} \}_{i \in \mathcal{L}_v \cup \mathcal{L}_t}$
denote the semantic feature selection strategy for SU-$k$, where $\beta_{k,i} \in \{0,1\}$ indicates whether the $i$-th semantic features is retained ($\beta_{k,i}=1$) or excluded  ($\beta_{k,i}=0$).
As such, the selected cross-modal features are obtained as follows:
\begin{equation}
(\mathbf{f}_{v,k}^{\text{selected}},\mathbf{f}_{t,k}^{\text{selected}}) = \boldsymbol{\beta}_k \odot (\mathbf{f}_{v,k},\mathbf{f}_{t,k}),
\end{equation}
where $\odot$ denotes the Hadamard product.
Thus, the traffic demands of the visual features $Q_{v,k}$ and the textual features $Q_{t,k}$ are represented as follows:
\begin{equation}
Q_{v,k} = \mathcal{C}_v(\mathbf{f}_{v,k}^{\text{selected}}) \text{ and } Q_{t,k} = \mathcal{C}_t(\mathbf{f}_{t,k}^{\text{selected}}),
\end{equation}
where $\mathcal{C}_v$ and $\mathcal{C}_t$ represent the channel encoding functions that map the cross-modal features into the bitstream.
\subsection{NOMA Transmission}
Semantic feature selection reshapes SUs' traffic demands, while NOMA allocates varying transmission capacities to them. This joint control significantly enhances flexibility and is expected to improve transmission efficiency. The AP with $Z$ receive antennas can introduce the receive beamforming to enhance the delivery of each SU's semantic features.  The receive beamforming strategy is defined as $\boldsymbol{\omega}=\{\boldsymbol{\omega}_k\}_{k\in\mathcal{K}}$~\cite{He-2023jsac}, where $\boldsymbol{\omega}_k$ denotes the receive beamforming assigned to SU-$k$.
NOMA allows multiple SUs to simultaneously access the wireless channel. Based on the NOMA decoding order, the superimposed signals are decoded by using the successive interference cancellation (SIC) method~\cite{Dai-2015ICM}. To characterize the NOMA decoding order, we introduce a binary variable $\pi_{k,j}$, where $\pi_{k,j} = 1$ indicates that SU-$k$ is decoded prior to SU-$j$, treating SU-$j$'s signal as interference. Thus, we have the following  NOMA decoding constraints for each SU:
\begin{subequations}\label{decoding-relation}
\begin{align}
&\pi_{k,j}+\pi_{j,k} = 1 \text{ and }
\pi_{k,j}\in\{0,1\},\label{ori-decoding}\\
&r_k < r_i + M (1-\pi_{k,j}),~ k\neq j, \forall k,j\in\mathcal{K},
\end{align}
\end{subequations}
where $\mathbf{r} = \{r_k\}_{k\in\mathcal{K}}$ denotes SUs' transmission priorities and $M$ is a sufficiently large constant to ensure acyclic decoding order~\cite{Huang-2024twc}.
Let $\mathbf{h}_k$ denote the wireless channel from SU-$k$ to the AP and $p_o$ denote the transmit power of SUs. Thus, the transmission capacity of SU-$k$ is expressed as follows:
\begin{equation}\label{transmission-capacity}
R_k \!=\! B\log_2\!\Big(\!1+\frac{|\mathbf{h}^H_k\boldsymbol{\omega}_k|^2p_o}{\sum\limits_{j\neq k,j\in\mathcal{K}}\!\!\!\!\!\!\pi_{k,j}|\mathbf{h}^H_j\boldsymbol{\omega}_k|^2 p_o\!+\!B\sigma^2}\!\Big),~\forall k \in \mathcal{K},
\end{equation}
where $B$ denotes the bandwidth and $\sigma^2$ denotes the noise power spectrum density. To ensure complete semantic information delivery, each SU is required to satisfy its traffic demand $R_k\ge Q_{v,k}+Q_{t,k}\triangleq Q_k$. Thus, the overall transmission latency $T$ is computed as follows:
\begin{equation}\label{latency}
T = \max_{k\in\mathcal{K}}\frac{Q_k}{R_k}.
\end{equation}

Note that the semantic feature selection strategy reshapes the traffic demands by controlling the size of the semantic features, while the AP's receive beamforming and the NOMA decoding order influence the transmission capacity by reconfiguring channel and interference conditions. Therefore, we aim to jointly consider the semantic and transmission controls to improve the semantic fidelity and transmission efficiency.

\subsection{GAI-assisted Semantic Image Recovery}
The AP employs a pre-trained SDM to recover the raw image from the received semantic features $\mathbf{f}_{k}=\{\mathbf{f}_{v,k}^{\text{selected}},\mathbf{f}_{t,k}^{\text{selected}}\}$, leveraging a generative denoising process to synthesize high-quality results~\cite{Liu-2024in}. Thus, the semantic feature selection directly influences the semantic recovery accuracy. The SDM can be replaced by other GAI-based models depending on the specific task requirements. Staring from a pure Gaussian noise $\mathbf{x}_{N,k} \sim \mathcal{N}(0, \mathbf{I})$, the SDM iteratively denoises it over $N$ steps to generate the recovered image $\mathbf{x}_{0,k}$ as follows:
\begin{equation}\label{denoising-1}
\mathbf{x}_{n-1,k} = \frac{1}{\sqrt{\alpha_n}} \left( \mathbf{x}_{n,k} - \frac{1 - \alpha_n}{\sqrt{1 - \bar{\alpha}_n}} \boldsymbol{\epsilon}_\theta(\mathbf{x}_{n,k}, n, \mathbf{f}_{k}) \right),
\end{equation}
where $\alpha_n$ is the controlling parameter determining the denoising strength at the $n$-th step and $\bar{\alpha}_n \triangleq \Pi_{i=1}^n \alpha_i$. Guided by $\mathbf{f}_{k}$, the denoising network  $\boldsymbol{\epsilon}_\theta$ predicts the noise and minimizes its deviation from the actual noise as follows:
\begin{equation}\label{denoising-2}
\mathcal{L}_\theta = \mathbb{E}_{\mathbf{x}_{0,k}, \mathbf{\epsilon}} \left\{ \left\| \boldsymbol{\epsilon}_\theta(\mathbf{x}_{n,k}, n, \mathbf{f}_{k}) - \boldsymbol{\epsilon}_{n,k} \right\|^2 \right\},
\end{equation}
where $\boldsymbol{\epsilon}_{n,k}$ is actual noise added to $\mathbf{x}_{n-1,k}$. We utilize the Learned Perceptual Image Patch Similarity (LPIPS) metric to evaluate the  recovery accuracy~\cite{Erdemir-2023jsac}, which measures perceptual similarity by comparing high-level feature representations between raw image $\mathbf{s}_k$ and recovered image $\mathbf{x}_{0,k}$ as follows:
\begin{equation}\label{semantic-similarity}
\text{LPIPS}_k = \sum_{l=1}^L w_l \cdot \left\| \mathcal{F}_l(\mathbf{s}_k) - \mathcal{F}_l(\mathbf{x}_{0,k}) \right\|_2^2,
\end{equation}
where $\mathcal{F}_l(\cdot)$ and $w_l$ represents the feature extractor and the associated weight at $l$-th layer of the evaluating deep network, such as VGG network~\cite{zhang2018unreasonable}. Parameter $L$ is the total number of network layers. Note that a smaller LPIPS value indicates a higher semantic recovery accuracy between $\mathbf{s}_k$ and  $\mathbf{x}_{0,k}$.  In~\eqref{denoising-1} and \eqref{denoising-2}, the encoded high-dimensional features of $\mathbf{f}_{k}$ progressively incorporate the semantic features in the denoising process, guiding the generation of the recovered image $\mathbf{x}_{0,k}$. This highlights that selecting semantic features with higher semantic information leads to better alignment of the recovered image $\mathbf{x}_{0,k}$ with the raw data $\mathbf{s}_k$. 
\subsection{Weighted Performance Maximization}
Delving into the GAI-based semantic NOMA transmissions, transmitting richer semantic features enables the AP to more accurately recover the raw image.  However, this also aggravates transmission delay due to increased traffic demands. Thus, we propose the JTSC scheme to balance the trade-off between the semantic control and transmission control processes. We minimize the weighted performance of LPIPS and transmission latency by jointly optimizing the NOMA decoding order $\boldsymbol{\pi} = \{\pi_{k,j}\}_{k,j\in\mathcal{K}}$, the AP's receive beamforming $\boldsymbol{\omega}=\{\boldsymbol{\omega}_k\}_{k\in\mathcal{K}}$, and the semantic feature selection strategy $\boldsymbol{\beta} = \{\boldsymbol{\beta}_k\}_{k\in\mathcal{K}}$ as follows:
\begin{subequations}\label{problem-formulation}
\begin{align}
\min_{\boldsymbol{\pi},\boldsymbol{\beta},\boldsymbol{\omega}}& \sum_{k\in\mathcal{K}}\text{LPIPS}_k + \psi T,\\
\mathrm {s.t.}
~&~ \eqref{semantic-extraction}-\eqref{semantic-similarity},
\end{align}
\end{subequations}
where $\psi$ denotes the weighting factor that adjusts the relative importance of the semantic recovery accuracy and the transmission latency.
Due to the strong variable coupling and incomplete model information, problem~\eqref{problem-formulation} is challenging for conventional optimization methods. Note that the semantic recovery relies on SDM, whose black-box nature makes it difficult to perform theoretical analysis or apply model-based optimization methods in a straightforward manner. While model-free DRL methods can learn all control variables via interaction with the environment, the high dimensionality of control variables poses significant challenges in achieving convergence to an optimal solution.
\section{Model-based Optimization for Transmission Control Variables}\label{opti-transmit}
\begin{figure}[t]
	\centering
	\includegraphics[width = 0.4\textwidth]{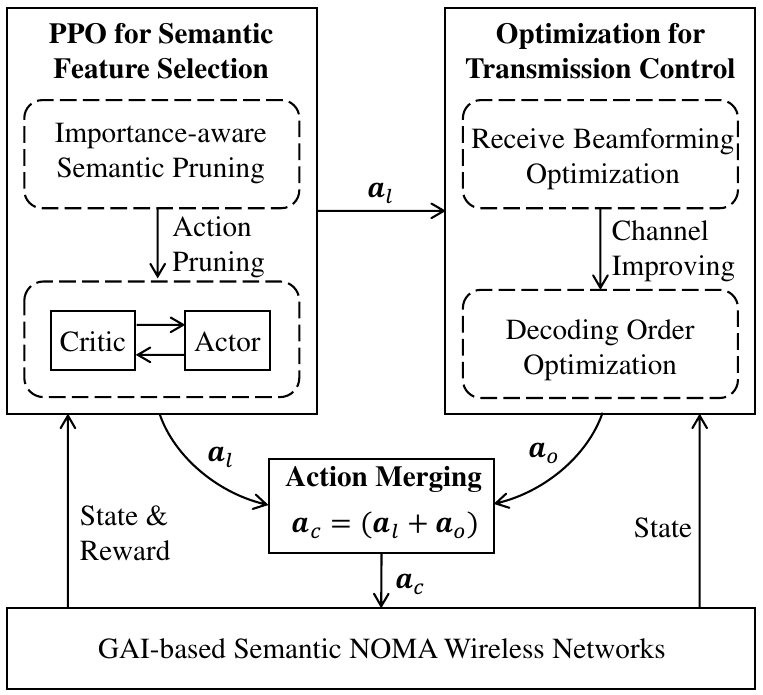}
	\caption{ Importance-aware and model-driven PPO framework.}\label{alg}
\end{figure}
One important insight is that incorporating model information into the learning methods can lead to more efficient learning efficiency~\cite{Shlezinger-2023}. This is because model information provides useful structural priors that help reduce the search space and guide the learning process toward more favorable regions. Motivated by this, we propose an IM-PPO framework to leverage advantages from both model-based optimizations and model-free PPO, as shown in Fig.~\ref{alg}. Specifically, we incorporate model information by two key aspects as follows:
\begin{itemize}
\item We decompose the control variables into the model-based and the model-free parts. The model-based part includes variables for which explicit model information is available, i.e., the receive beamforming and the NOMA decoding order, and thus can be efficiently optimized using model-based optimization methods. The model-free part consists of variables associated with the black-box component, i.e., the semantic feature selection, which is learned through the model-free PPO method.
\item Although the SDM is based on a GAI architecture whose internal logic is complex to analyze, we propose a cross-modal matching approach to reveal the underlying interplay between textual features and visual features. This allows us to evaluate the importance of semantic features and prune low-importance ones, thus significantly enhancing PPO's learning efficiency.
\end{itemize}

In the sequel, we detail the IM-PPO framework for optimizing the transmission control variables and the semantic feature selection in Section~\ref{opti-transmit} and Section~\ref{impor-ppo}, respectively. We first design model-based optimization methods for the transmission control variables in the following.
\subsection{Receive Beamforming Optimization}\label{receive-beam}
The receive beamforming helps refine the channel conditions for SUs, thereby improving the NOMA decoding flexibility at the AP. We first optimize the receive beamforming by solving the following subproblem for each SU:
\begin{equation}\label{receive-beamforming}
\max_{\boldsymbol{\omega}_k} \frac{\boldsymbol{\omega}^H_k \mathbf{H}_k\boldsymbol{\omega}_k}{\boldsymbol{\omega}^H_k \overline{\mathbf{H}}_k\boldsymbol{\omega}_k},~\forall k\in\mathcal{K},
\end{equation}
where $\mathbf{H}_k \triangleq\mathbf{h}_k\mathbf{h}^H_kp_o$ and $\overline{\mathbf{H}}_k \triangleq \!\!\!\sum\limits_{j\neq k,j\in\mathcal{K}}\!\!\!\!\!\!\pi_{k,j}\mathbf{h}_j\mathbf{h}^H_jp_o+B\sigma^2\mathbf{I}$.
Problem~\eqref{receive-beamforming} is a standard generalized eigenvalue problem~\cite{Hu-2021tcom}, where the optimal receive beamforming is obtained by the eigenvector of the largest eigenvalue as follows:
\begin{equation}
\boldsymbol{\omega}^*_k = \text{eigvec}_{\max} \left\{ \overline{\mathbf{H}}_k^{-1} \mathbf{H}_k \right\},~\forall k \in \mathcal{K},
\end{equation}
where $\text{eigvec}_{\max}\{\cdot\}$ denotes the eigenvector corresponding to the largest eigenvalue. To avoid repeatedly updating the receive beamforming under varying NOMA decoding order, we consider a worst-case decoding condition where all SUs are decoded first, similar to that in~\cite{Zhao-2025twc}.
\subsection{NOMA Decoding Order Optimization}\label{decoding-order}
The NOMA decoding order determines the transmission priority of SUs, which affects their individual transmission capacities. Therefore, optimizing the NOMA decoding order aims to ensure that each SU's traffic demand is effectively satisfied. By treating the latency $T$ as an auxiliary variable, we formulate the NOMA decoding order optimization as follows:
\begin{subequations}\label{subproblem-decoding}
\begin{align}
\min_{{\boldsymbol \pi},T}&~T, \label{obj-decoding}\\
\mathrm {s.t.}
~& \eqref{decoding-relation}\text{ and }B\log_2\left(1+\gamma_k\right)\ge \frac{Q_k}{T},~\forall k \in\mathcal{K},\label{re-sinr}
\end{align}
\end{subequations}
where $\gamma_k$ is the signal-to-interference-plus-noise ratio (SINR) of SU-$k$ as represented in~\eqref{transmission-capacity}. To proceed, we employ quadratic transform~\cite{Shen-2018itsp} to approximate  $\gamma_k$ as follows:
\begin{equation}\label{SINR-QT}
\gamma_k \!\leq\! 2y_k\!\sqrt{|\mathbf{h}^H_k\boldsymbol{\omega}_k|^2p_o}\!-\!y_k^2(\!\!\!\!\sum\limits_{j\neq k,j\in\mathcal{K}}\!\!\!\!\pi_{k,j}|\mathbf{h}^H_j\boldsymbol{\omega}_k|^2 p_o\!+B\sigma^2),
\end{equation}
where ${\bf y}=\{ y_k \}_{k\in\mathcal{K}}$ is an auxiliary variable. Note that constraint~\eqref{SINR-QT} is convex with respect to $y_k$. The SINR~\eqref{SINR-QT} remains equivalent to that in~\eqref{transmission-capacity} when the auxiliary variable $y_k$ satisfies the following condition:
\begin{equation}\label{obtain-y}
y_k^*= \frac{\sqrt{|\mathbf{h}^H_k\boldsymbol{\omega}_k|^2p_o}}{\!\!\!\sum\limits_{j\neq k,j\in\mathcal{K}}\!\!\pi_{k,j}|\mathbf{h}^H_j\boldsymbol{\omega}_k|^2 p_o+B\sigma^2},~k\in\mathcal{K}.
\end{equation}
Therefore, we can optimize problem~\eqref{subproblem-decoding} by iteratively updating $y_k$.  However, constraint~\eqref{decoding-relation} remains non-convex due to the integer nature. To tackle this, we reformulate constraint~\eqref{ori-decoding} into a continuous form as follows:
\begin{subequations}\label{decoding-relaxation}
\begin{align}
&0 \leq \pi_{k,j} \!\leq\! 1 \text{ and }\pi_{k,j}\!+\!\pi_{j,k} - 1 =0,   \label{binary}    \\
&\pi_{k,j}-\pi_{k,j}^2 = 0 ,~k\neq j,~\forall k,j\in\mathcal{K}.\label{taylor}
\end{align}
\end{subequations}
By employing the first-order Taylor expansion, constraints~\eqref{taylor} can be linearly approximated as follows:
\begin{equation}\label{tylor-pi}
\pi_{k,j}+\pi^2_{k,j,0}-2 \pi_{k,j} \pi^2_{k,j,0} \le 0,~\forall k,j \in \mathcal{K},
\end{equation}
where $\pi_{k,j,0}$ denotes the expansion point, which is updated at each iteration.
By incorporating constraint~\eqref{tylor-pi} into problem~\eqref{subproblem-decoding}, it becomes a convex optimization problem that can be efficiently solved using standard convex tools.
\section{Importance-aware DRL for Semantic Feature Selection}\label{impor-ppo}
After obtaining the transmission control variables, a straightforward approach is to employ DRL to select the most informative semantic features from the original feature space. However, as the feature space grows, this leads to significant training overhead and increases the risk of the DRL algorithm converging to local optima. This is mainly because the model-free DRL does not exploit underlying model information, and thus its performance heavily depends on the quality of experience gathered through environment interactions. To address this challenge, we propose a cross-model matching method to introduce model knowledge into the DRL framework. As such, the DRL agents can reduce the reliance on trial-and-error exploration, thus improving the learning efficiency significantly.
\subsection{Importance-aware Semantic Feature Pruning}
We leverage the pre-trained BLIP model to generate textual features. However, the BLIP model may generate redundant textual features due to overfitting to raw images. To reduce redundancy, we identify the attribution of each textual feature element to the raw images using the Grad-CAM method~\cite{Liu-2024-itac}, and select  the most informative textual features accordingly. As shown in Fig.~\ref{cross-modal-promt-refine}, the raw image is first processed through an image encoder to extract visual high-level features $\mathbf{A}= \{A_m\}_{m\in\mathcal{M}}$, where $\mathcal{M}$ denotes the set of different image regions. These features capture the semantic information from various regions of the raw image, which guides  the text decoder in generating relevant textual features.
We compute the gradient of each textual feature element $y_i$  to $\mathbf{A}$, thereby quantifying the contribution $\alpha_{i,m}$ of the $i$ textual feature element to the $m$-th image region as follows:
\begin{equation}\label{sum-grad}
\alpha_{i,m} = \frac{\partial y_i}{ \partial A_m}, ~ \forall m \in \mathcal{M}.
\end{equation}
Therefore, we generate a heatmap $\mathbf{H} =\{H_i\}_{i\in\mathcal{L}_t}$, highlighting the important regions in the raw image that contribute to generating the corresponding textual feature, as follows:
\begin{equation}\label{heatmap}
H_i = \text{ReLU}\big(\sum_{m\in\mathcal{M}}\alpha_{i,m}A_m\big),~\forall i\in\mathcal{L}_t,
\end{equation}
where $\text{ReLU}(\cdot)$ is the linear rectification function, as we only focus on identifying the high-level features that positively influence the generative textual features. Given $\mathbf{H}$, we can evaluate the importance of the textual features elements by exploring their dependency and contribution as follows:
\begin{figure}[t]
	\centering
	\includegraphics[width = 0.45\textwidth]{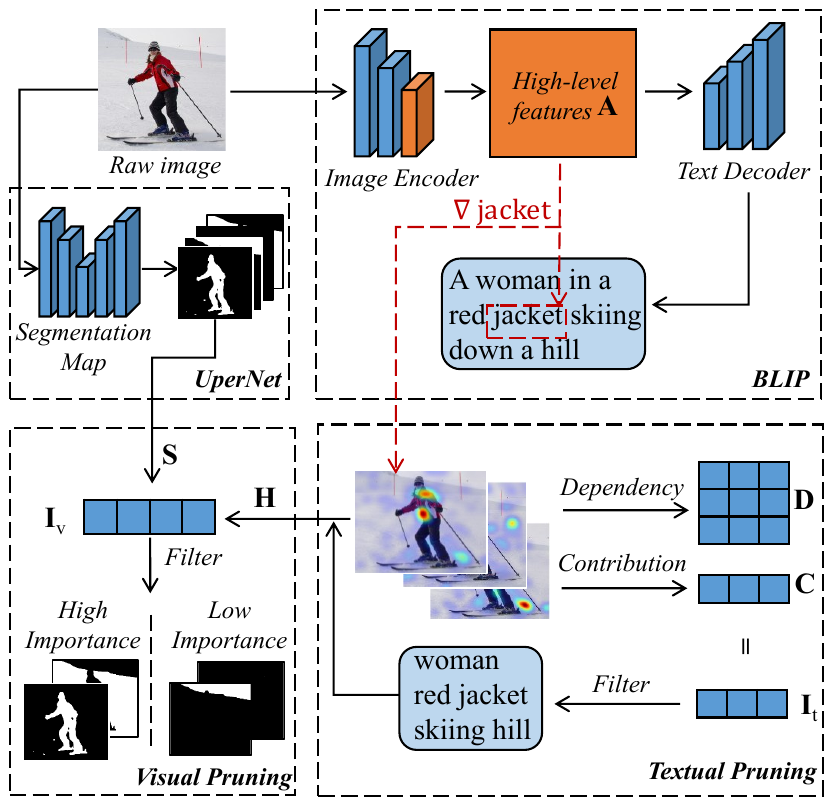}
	\caption{Importance-aware semantic feature pruning.}\label{cross-modal-promt-refine}
\end{figure}
\subsubsection{Dependency}
The dependency refers to the degree of mutual influence between two elements. For example, as shown in Fig.~\ref{cross-modal-promt-refine}, the heatmaps of the textual features \texttt{[woman]} and \texttt{[jacket]} overlap significantly, indicating a strong correlation between them. This is reasonable because a woman typically wears a jacket. However, the overlap between \texttt{[jacket]} and \texttt{[hill]} is smaller, suggesting a weaker correlation between them. Therefore, we can design the dependency matrix $\mathbf{D}=\{D_{i,j}\}_{i,j\in\mathcal{L}_t}$ to represent the dependency between any two elements as follows:
\begin{equation}
D_{i,j}= \frac{|H_i \cap H_j|}{|H_i \cup H_j|},~\forall i,j\in\mathcal{L}_t.
\end{equation}
\subsubsection{Contribution}
The contribution reflects how much each semantic feature contributes to the raw image. The region activated in the heatmap reflects the spatial impact of the corresponding textual feature on the raw image. A larger activated region implies that the associated textual feature contributes more significantly to the raw image. Therefore, we define the contribution matrix $\mathbf{C}=\{C_i\}_{i\in\mathcal{L}_t}$ as follows:
\begin{equation}
C_i = \frac{|H_i|}{\sum_{j\in\mathcal{L}_t}|H_j|},~\forall i\in\mathcal{L}_t.
\end{equation}
To assess the importance of textual features, we define the textual importance score $\mathbf{I}_t=\{I_{t,i}\}_{i\in\mathcal{L}_t}$ by jointly considering the impacts of both contribution and dependency as follows:
\begin{equation}
I_{t,i}= \sum_{j\in\mathcal{L}_t}C_iD_{i,j},~\forall i\in\mathcal{L}_t.
\end{equation}
The text feature elements with higher $I_{t,i}$ values are considered more semantically influential.  As such, we introduce a predefined threshold $\xi_t$ and retain only those features satisfying $I_{t,i} \ge \xi_t$ to filter out those with low semantic importance.

We employ the UperNet model to extract semantic segmentation maps as visual features. However, this process can also segment out semantically irrelevant or less important categories, such as background regions, which increases both the transmission overhead and the computational burden during the image recovery. Therefore, we design a cross-modal matching method to align the visual features with the filtered textual features to enable more accurate image generation.
We exploit the spatial correspondence between the textual heatmaps and the visual segmentation maps. Intuitively, a segmentation category is considered more informative if its region overlaps with the activation areas of multiple textual heatmaps~\cite{Farrag-2023gc}.  Let $\mathbf{S} = \{S_n\}_{n\in\mathcal{N}}$ denote the set of segmentation masks, where $S_n$ represents the region of the $n$-th segmented category. Hence, we define an importance matrix $\mathbf{I}_v =\{I_{v,n}\}_{n\in\mathcal{L}_v}$ to quantify the importance of each visual feature with the filtered textual features as follows:
\begin{equation}\label{importance-index}
I_{v,n}= \sum_{i \in \mathcal{L}_t} \frac{|S_n \cap H_i|}{|S_n|},~\forall n \in \mathcal{L}_v.
\end{equation}
A higher value of $I_{v,n}$ indicates that the segmented region $S_n$ carries more semantic information. Similarly, we introduce a threshold $\xi_v$ to retain the segmentation categories with $I_{v,n} \geq \xi_v$, thus preserving visual features with higher semantic information. Both $\xi_t$ and $\xi_v$ can be pre-obtained depending on the requirements of the target applications.
\subsection{PPO for Semantic Feature Selection}
After importance-aware semantic feature refinement, the filtered cross-modal features exhibit higher semantic information. However, it is not necessary to transmit all of these features due to the following reasons:
\begin{itemize}
\item \emph{Uncertainty in Transmission Capacities}:
Due to the dynamic transmission capacities and latency requirements of SUs, it may be infeasible to transmit all the filtered semantic features. Hence, a further semantic feature selection is required to effectively reduce the traffic demands and ensure communication reliability.
\item \emph{Redundancy across Cross-modal Features}:
The semantic information conveyed by textual and visual features is often complementary but can also be partially redundant. For example, if a textual feature sufficiently captures a specific semantic concept, its semantically related visual feature may offer redundant information. As such, by identifying and eliminating such cross-modal redundancy, the transmission efficiency can be further improved.
\end{itemize}

To this end, the PPO method is employed to learn the semantic feature selection strategy, as its advantage estimation and clipping mechanism can effectively stabilize training and improve learning efficiency~\cite{Zhang-2023jsac}. As shown in Fig.~\ref{alg}, the IM-PPO integrates both the model-based optimization and the model-free PPO modules. The PPO is characterized by the tuple $(\mathcal{S},(\mathcal{A}_l,\mathcal{A}_o),\mathcal{R})$. The state $s\in\mathcal{S}$ includes all SUs' channel conditions and the NOMA decoding order. Depending on the availability of analytical model information, we decompose all control variables into two action spaces, i.e., $\mathcal{A}_l$ and $\mathcal{A}_o$, where $\mathbf{a}_l \in \mathcal{A}_l$ represents the semantic feature selection strategy $\boldsymbol{\beta}$ and $\mathbf{a}_o \in \mathcal{A}_o$ includes the receive beamforming strategy $\boldsymbol{\omega}$ and the NOMA decoding order $\boldsymbol{\pi}$. Before the PPO agent selects semantic features, we perform the cross-modal matching method to identify those with high semantic importance. The importance-aware feature pruning effectively reduces the dimensionality of the action space $\mathcal{A}_l$ and improves exploration efficiency. Based on the SUs' transmission capacities, the PPO selects the semantic features to contain more semantic information while satisfying the SUs' traffic demands.  Given $\mathbf{a}_l$ from the PPO module, the model-based optimization optimizes $\mathbf{a}_o$ for efficient NOMA transmission. By merging the actions from both modules $\mathbf{a}_c=(\mathbf{a}_l+\mathbf{a}_o)$ and executing it into the wireless network, we obtain the next state and the feedback reward $r\in\mathcal{R}$, which update the PPO's policy in the next learning round.

To consider both the semantic fidelity and transmission efficiency, we define the reward function as follows:
\begin{equation}\label{reward}
r = -\Big(\sum_{k\in\mathcal{K}}\text{LPIPS}_k + \psi T\Big)-\mathcal{P}(\mathbf{a}_c),
\end{equation}
where $\mathcal{P}(\mathbf{a}_c)$ represents a penalty function that imposes a sufficiently large cost when the constraints in problem~\eqref{problem-formulation} are violated. The entire procedure of the IM-PPO framework is summarized in Algorithm~\ref{alg-mkppo}. As in lines~\ref{action-pruning}--\ref{semantic}, the importance-aware semantic feature refinement first filters the features with high semantic importance, after which the PPO agent adapts the semantic feature selection strategy within reduced exploration space. Given the selected semantic features, the SUs' traffic demands can be determined, allowing the optimization module to solve the transmission control variables, as in line~\ref{optimize-subproblem}. In lines~\ref{com-act}--\ref{update-networks}, the PPO agent merges the actions from both the PPO and optimization modules and executes them in the environment, such that the two modules remain synchronized in each learning round. The resulting feedback reward and next state are stored in the memory buffer and used to update the PPO's decision-making policy. The computational complexity of Algorithm~\ref{alg-mkppo} is computed as $C_p+C_o$, where $C_p$ and $C_o$ represent the complexities of the PPO and optimization modules, respectively. We define $ n_{a,f}$ and $n_{c,f}$ as  the number of neurons in the $f$-th layer of the actor and critic networks, respectively. Thus, the complexity of PPO module is represented as $C_p = \mathcal{O}\big(\sum_{f=0}^{F_a-1}n_{a,f}n_{a,f+1} + \sum_{f=0}^{F_c-1}n_{c,f}n_{c,f+1}\big)$~\cite{Guo-ton2023}, where $F_a$ and $F_c$ denote the total number of layers in the actor and critic networks, respectively. The optimization module introduces a complexity of $C_o = \mathcal{O}\big((K^2+1)^{3.5}+K\big)$~\cite{Luo-2020spm}.
\begin{algorithm}[t]
	\caption{IM-PPO for Semantic Fidelity and Transmission Efficiency in GAI-based Semantic Image Transmissions}\label{alg-mkppo}
	\begin{algorithmic}[1]
        \State Initialize DNN parameters and all control variables.
        \Statex \textbf{\% IM-PPO for Semantic Feature Selection}
        \State \hspace{3mm} Perform importance-aware action pruning \label{action-pruning} via~\eqref{sum-grad}-\eqref{importance-index}\label{Cross-modal}
        \State \hspace{3mm} Observe the current system state $s$\label{observe-state}
        \State \hspace{3mm} PPO adapts the semantic features strategy $\boldsymbol{\beta}$ \label{semantic}
        \Statex \textbf{\% Optimization for Transmission Control}
        \State \hspace{3mm} Given~$\boldsymbol{\beta}$,  optimize $\boldsymbol{\omega}$ and $\boldsymbol{\pi}$ by solving~\eqref{receive-beamforming} and~\eqref{subproblem-decoding} \label{optimize-subproblem}
        \State \hspace{3mm}  Execute the merged action $\mathbf{a}_{c}=(\mathbf{a}_{l},\mathbf{a}_{o})$\label{com-act}
        \State \hspace{3mm} Receive the  feedback reward $r$ \label{observe-reward}
        \State \hspace{3mm} Record the transition to the next state ${\bf s}'$ \label{next-state}
        \State \hspace{3mm} Store the transition tuple $\{{\bf s},\mathbf{a}_{c}, r, {\bf s}'\}$ into  memory\label{store-transition}
        \State \hspace{3mm} Sample from memory and update PPO parameters  \label{update-networks}
	\end{algorithmic}
\end{algorithm}
\begin{figure*}[t]
	\centering 
\subfloat[Convergence performance.]{\includegraphics[width=0.322\textwidth]{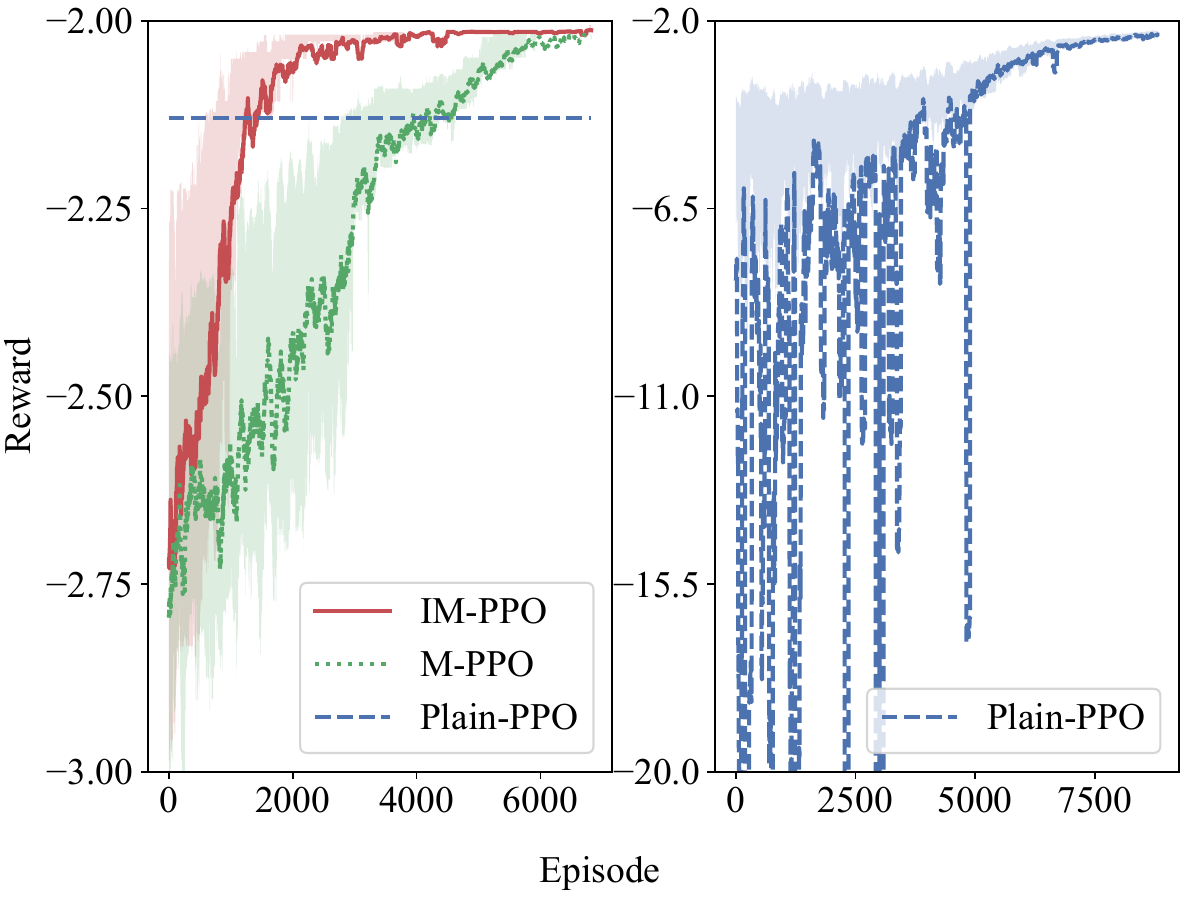}}
\subfloat[Latency and LPIPS.]{\includegraphics[width=0.33\textwidth]{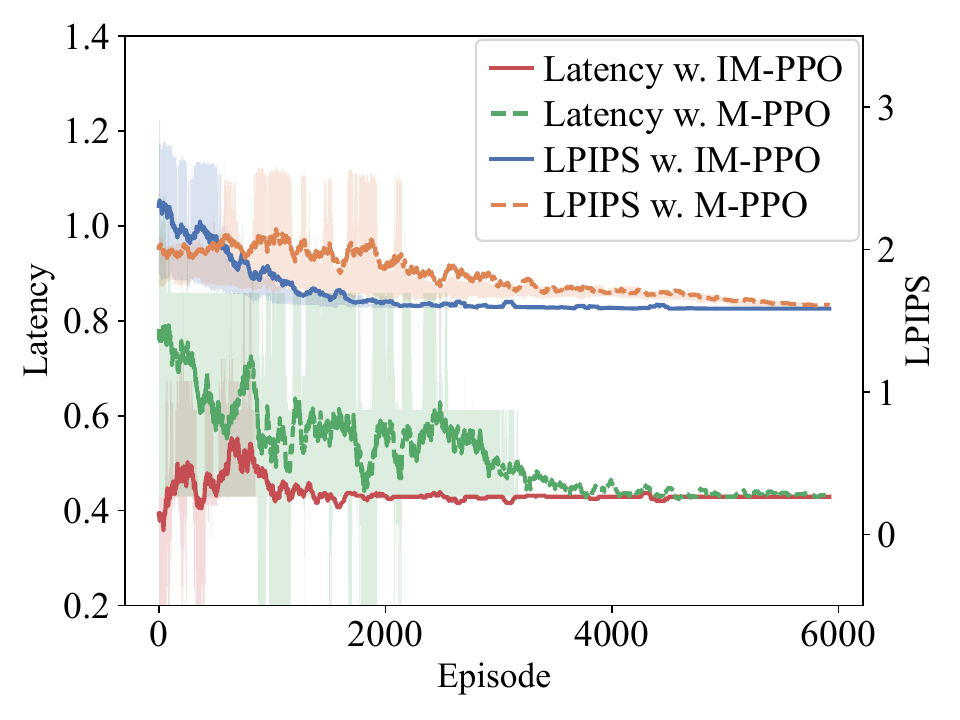}}
\subfloat[Feature selection ratio.]{\includegraphics[width=0.33\textwidth]{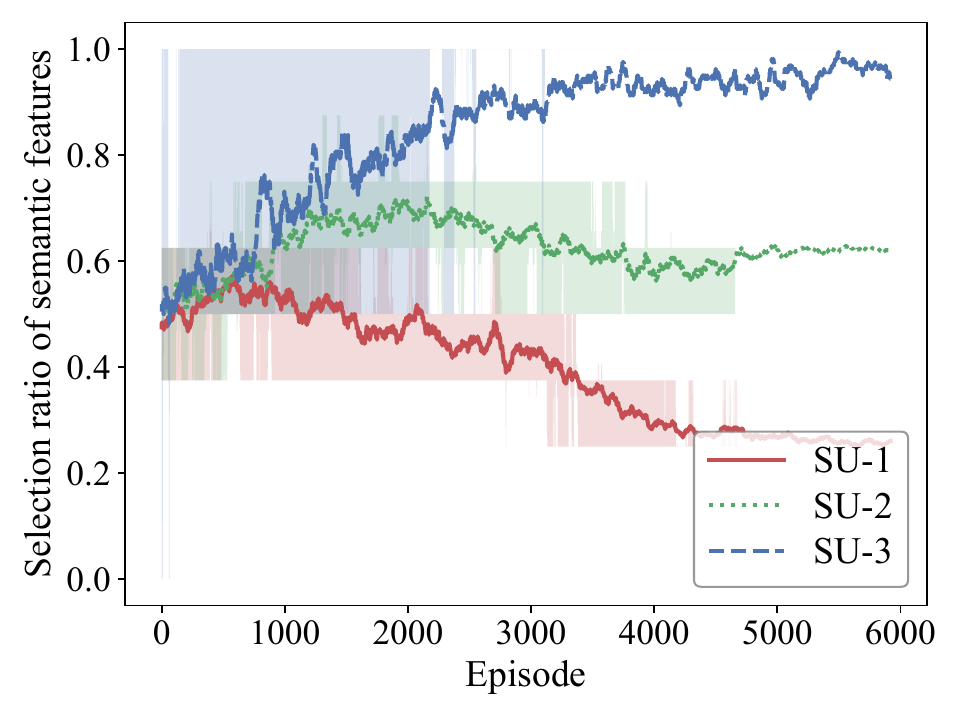}}
\caption{Learning performance of the IM-PPO framework.}\label{algorithm}
\end{figure*}
\begin{figure*}[t]
	\centering \subfloat[Different bandwidths.]{\includegraphics[width=0.32\textwidth]{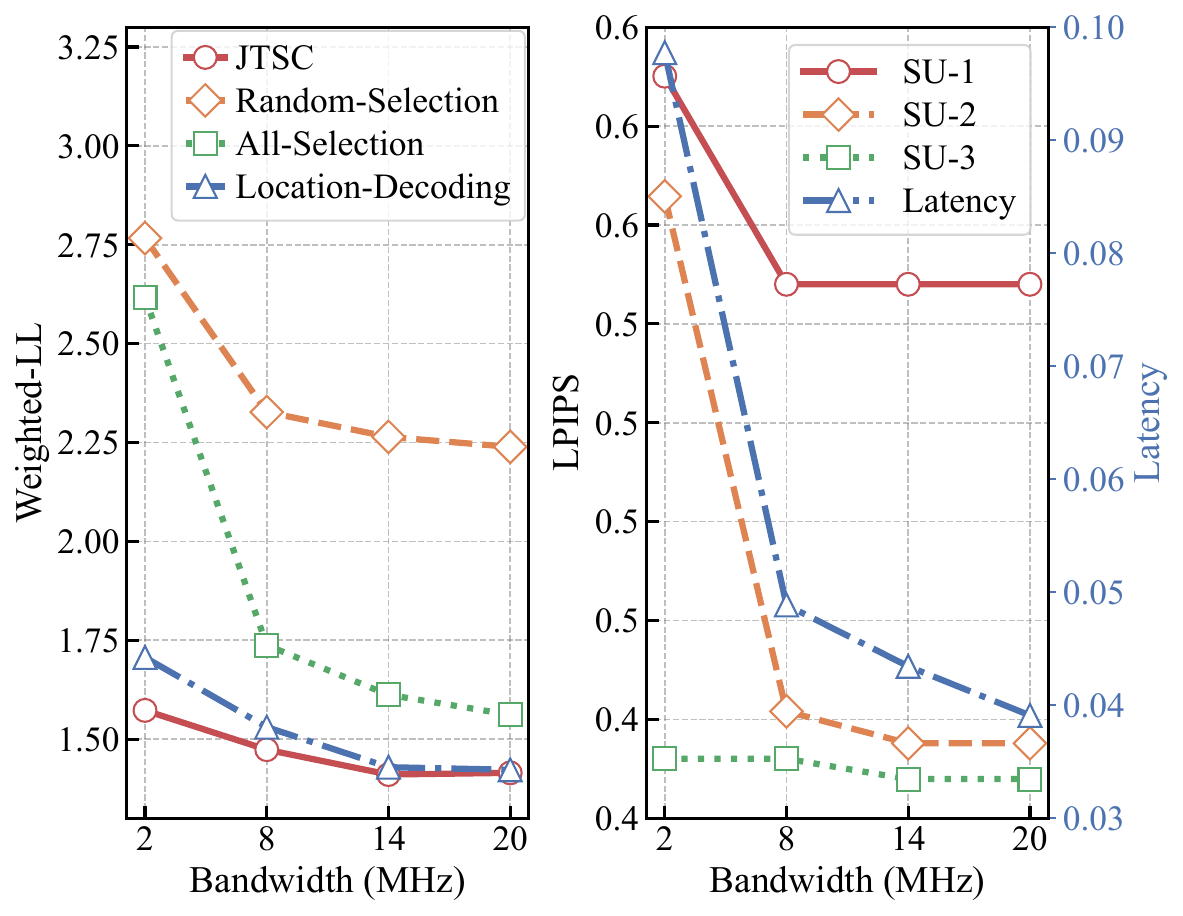}}
\subfloat[Different numbers of SUs.]{\includegraphics[width=0.32\textwidth]{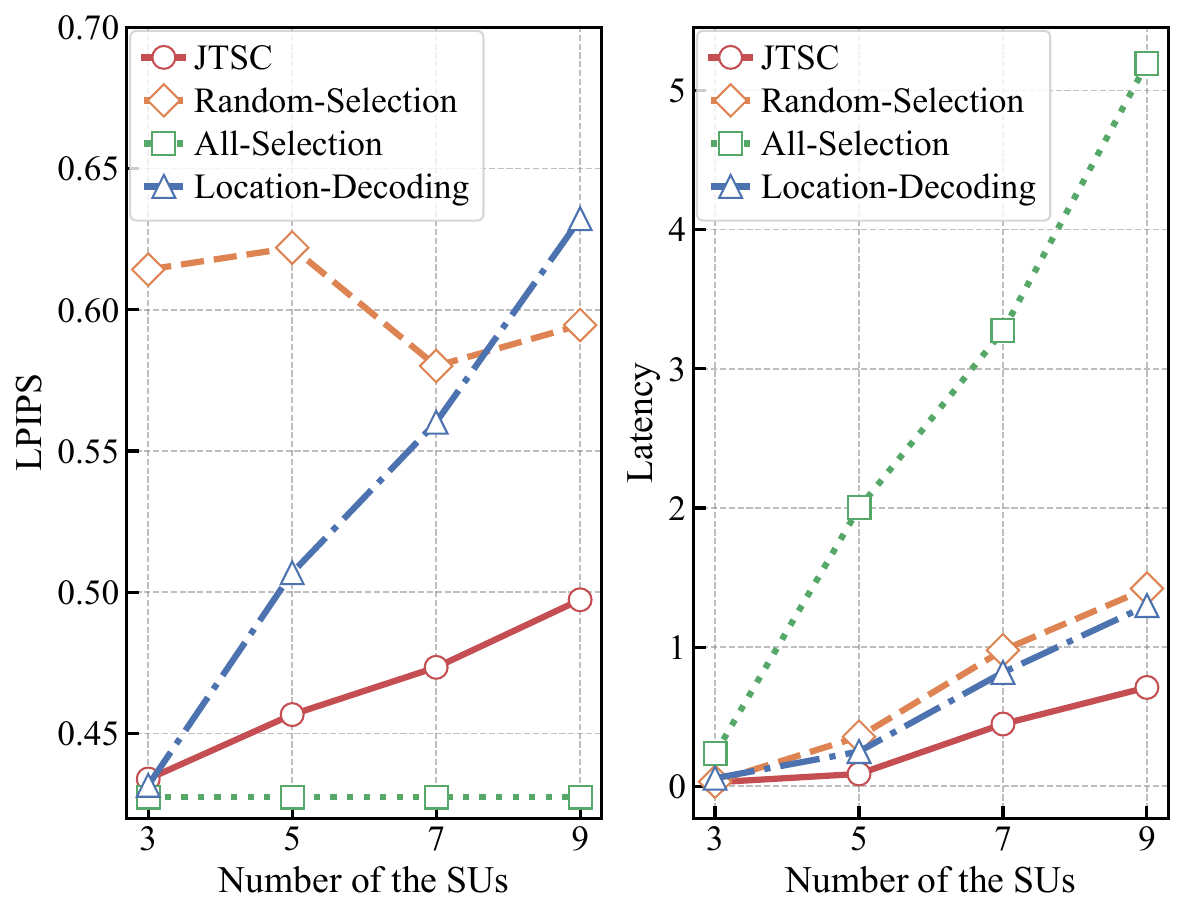}}
\subfloat[Different power of SU-3.]{\includegraphics[width=0.32\textwidth]{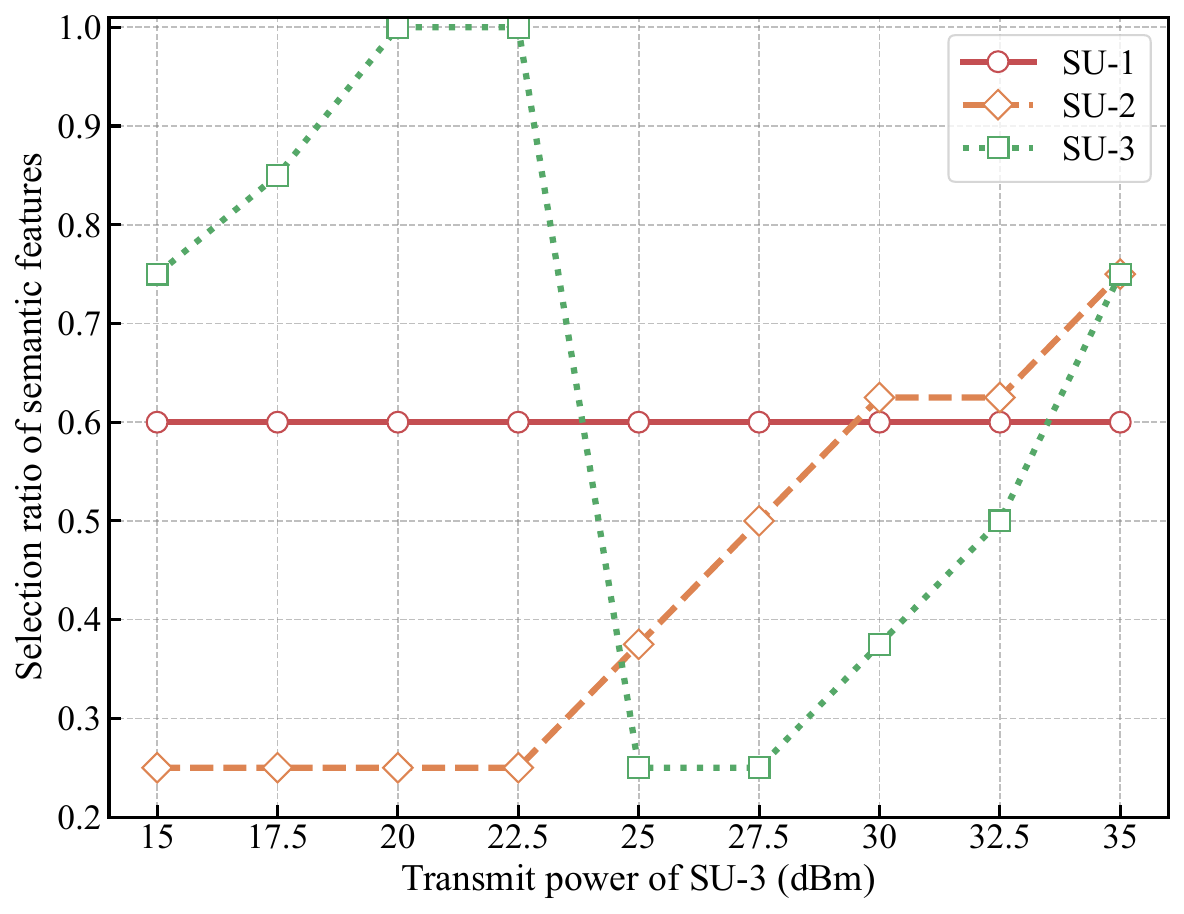}}
\caption{System performance enhancement by JTSC scheme.}\label{Weighted-per}
\end{figure*}
\section{Numerical Results}\label{results}
In this section, we evaluate the JTSC scheme in GAI-assisted semantic NOMA transmissions using the proposed IM-PPO framework. Stable Diffusion v1-5~\cite{stablediffusionmodel2023} is adopted as the backbone of the SDM model at the AP side. The raw images used for semantic transmission are obtained from the COCO-2017 dataset~\cite{COCO-2017dataset}, which is widely adopted in vision and semantic communication research. Following a similar setup in~\cite{Zhao-2023iotj}, the learning rates of the actor and critic networks are set to $1 \times 10^{-4}$. The default settings of the remaining simulation parameters are summarized as follows: number of SUs $K = 3$, SUs' transmit power $p_o = 30$ dBm,  number of AP's antennas $Z = 4$, bandwidth $B=2$ MHz, noise power spectrum density $\sigma^2 = -174$ dBm/Hz, number of denoising steps $N = 10$, and the filter thresholds for textual and visual features $\xi_t = 0.5$ and $\xi_v = 0.1$, respectively.
\subsection{Convergence of IM-PPO Framework}
To validate the convergence of the IM-PPO framework, we evaluate its learning performance compared to two benchmark methods:  M-PPO~\cite{Zhao-2023iotj} and Plain-PPO methods. Specifically, the M-PPO method integrates the model-based optimization module but does not apply cross-modal matching for semantic feature pruning. The Plain-PPO method adopts a model-free PPO approach to directly optimize all control variables without leveraging any model information.

Figure~\ref{algorithm}(a) studies the convergence performance of different algorithms. It is observed that the proposed IM-PPO achieves the fastest convergence, requiring approximately $2,500$ episodes, compared to around $6,000$ episodes for M-PPO and over $7,500$ episodes for Plain-PPO. These results demonstrate that incorporating model information effectively reduces inefficient exploration, thus guiding the DRL agent toward the optimal solution more efficiently. Moreover, the cross-modal matching method significantly reduces the action space for the PPO agent, which further accelerates its learning efficiency. In the left subfigure of Fig.~\ref{algorithm}(a), the blue dashed line indicates the converged average reward of Plain-PPO. We observe that both IM-PPO and M-PPO achieve higher final rewards, which indicates that integrating model information helps avoid local optima during the learning process.

Besides the overall reward performance, we further investigate the learning performance of individual metrics, i.e., LPIPS and transmission latency, as achieved by IM-PPO and M-PPO. As shown in Fig.~\ref{algorithm}(b), both methods successfully converge on each metric, which validates that the designed weighted reward function in~\eqref{reward} can effectively guide the learning of distinct objectives. Interestingly, although IM-PPO prunes a subset of semantic features through cross-modal matching, it still achieves a recovery accuracy comparable to that of M-PPO. This validates that the proposed cross-modal matching approach  effectively preserves the most informative semantic features.
To further analyze the interplay between the semantic control and the transmission capacity, we examine the semantic feature selection ratio of each SU, as shown in Fig.~\ref{algorithm}(c). We consider a fair setup where each SU transmits the same image. After executing the IM-PPO framework, the NOMA decoding order is assigned as SU-$1$, SU-$2$, and SU-$3$. We observe that the earlier-decoded SUs tend to select fewer semantic features to align with their transmission capacities. This verifies the effectiveness of IM-PPO in jointly optimizing the semantic feature selection and transmission control.
\subsection{Performance Enhancement by the JTSC Scheme}
This section studies the system performance improvements achieved by the proposed JTSC scheme. To highlight its advantages, we compare JTSC with three benchmark schemes, i.e., ALL-Selection~\cite{Du-2024ICASSP}, Random-Selection, and Location-Decoding~\cite{Shang-2023iotj}. In ALL-Selection, all semantic features are transmitted. The Random-Selection scheme transmits a randomly selected subset of semantic features, without considering their importance. In Location-Decoding, the NOMA decoding order is determined solely based on the distances between the SUs and the AP, and is not jointly optimized with the semantic feature selection strategy.

Figure~\ref{Weighted-per}(a) evaluates the system performance under different transmission capacities, where a higher bandwidth indicates stronger transmission capabilities for each SU. We define Weighted-LL as a composite metric to reflect both LPIPS and transmission latency performance, where lower values indicate better overall performance. In the left subfigure of Fig.~\ref{Weighted-per}(a), the Weighted-LL performance of all schemes improves as the bandwidth increases. It is observed that the JTSC scheme achieves the best Weighted-LL performance. The reason is that the JTSC scheme adaptively aligns  the semantic feature selection strategy to the SUs' transmission capacities. In contrast, the Random-Selection scheme randomly selects semantic features, leading to the loss of critical semantic information. Although ALL-Selection ensures high semantic fidelity by transmitting all features, it introduces significant transmission latency,  resulting in suboptimal overall performance. Location-Decoding fails to adjust the SUs' NOMA decoding order according to the current semantic feature selection strategy, also leading to increased transmission latency. As the SUs' transmission capacities increase, the performance gap from the JTSC scheme to the ALL-Selection and Location-Decoding schemes decreases due to the alleviation of transmission bottlenecks. The right subfigure of Fig.~\ref{Weighted-per}(a) evaluates the overall latency and the individual LPIPS.  As the SUs' transmission capacities increase, each SU transmits richer semantic features, thus improving the semantic recovery accuracy. Meanwhile, the latency also decreases due to the improvement in transmission capacities.
\begin{figure}[t]
	\centering
	\includegraphics[width = 0.45\textwidth]{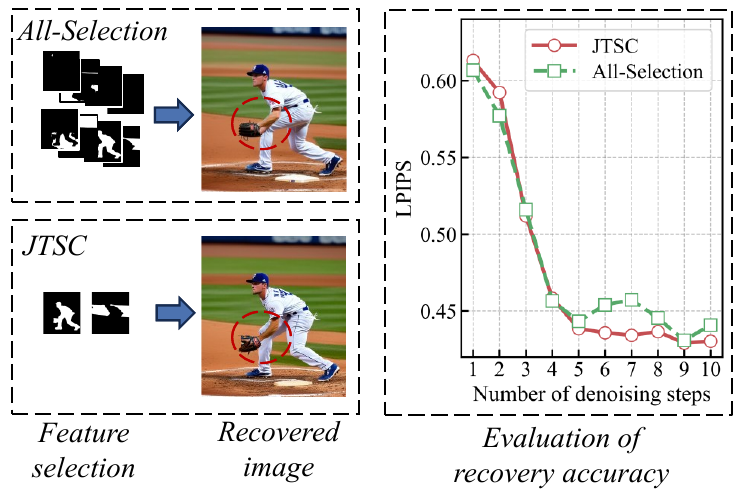}
	\caption{Visual validation of semantic feature selection.}\label{visual-show}
\end{figure}

In Fig.~\ref{Weighted-per}(b), we further investigate the LPIPS and latency performance under different numbers of SUs. As the number of SUs increases, the access competition among SUs becomes more severe.  In the left subfigure of Fig.~\ref{Weighted-per}(b), the ALL-Selection scheme consistently achieves the highest semantic recovery accuracy across different numbers of SUs. However, as shown in the right subfigure of Fig.~\ref{Weighted-per}(b), this comes at the cost of significantly increased transmission latency, which renders it impractical for real-time transmission scenarios.
Conversely, the JTSC scheme can balance LPIPS and latency performance, ensuring that both metrics degrade moderately as access competition continuously intensifies. This complementary design allows JTSC to effectively adapt  to varying transmission conditions, thereby enhancing the overall robustness. Note that the LPIPS performance of the Random-Selection scheme does not exhibit a clear trend as the number of SUs increases due to the lack of guidance from model information.

To investigate inter-SU impacts, we gradually increase the transmit power of SU-$3$ and compare the semantic feature selection ratios among the three SUs, as shown in Fig.~\ref{Weighted-per}(c). When the transmit power of SU-$3$ is low (below $22.5$ dBm), the optimized NOMA decoding order is SU-$2$, SU-$1$, and SU-$3$.
In this case, as SU-$3$'s transmit power increases, its transmission capacity also improves, enabling it to select more semantic features to enhance its recovery accuracy. Meanwhile, the semantic selection strategies of SU-$2$ remain unchanged due to the intensified interference from SU-$3$, forcing it to select only the most essential semantic features to ensure reliable semantic recovery.
When SU-$3$'s transmit power exceeds $25$ dBm, its interference to the earlier-decoded SU-$1$ and SU-$2$ becomes significant, making the current NOMA decoding order unsuitable. As such, the JTSC scheme adaptively reassigns the NOMA decoding order to SU-$3$, SU-$2$, and SU-$1$. After moving to the first decoding position, SU-$3$ has a significant reduction in transmission capacity, forcing it to reduce the number of selected semantic features to maintain effective adaptation. As SU-$3$'s transmit power continues to increase, its transmission capacity gradually improves. Meanwhile, SU-$2$ begins to increase its semantic feature selection ratio to improve recovery accuracy. This is because the increase in SU-$3$'s transmission capacity effectively relaxes the overall latency requirements, allowing SU-$2$ to transmit more semantic features without causing excessive delay. We observe that SU-$1$'s semantic feature selection ratio remains unchanged throughout the process, as its optimal semantic selection strategy always aligns with its transmission capacity.
\subsection{Visual Evaluation of Semantic Feature Selection}
Figure~\ref{visual-show} visually evaluates the impacts of semantic feature selection on semantic recovery accuracy. The ALL-Selection scheme selects all semantic features, while JTSC adaptively selects only the most informative ones.
Interestingly, as the number of denoising steps increases, JTSC gradually outperforms ALL-Selection in semantic recovery accuracy, and achieves more stable performance beyond five denoising steps. The rationale behind this result is that transmitting excessive semantic features may introduce redundant or conflicting information into the recovery process, thus weakening its ability to accurately capture and recover the critical semantic details.
In contrast, transmitting fewer but more informative semantic features allows the generative process to focus on the most essential features, resulting in more precise and faithful semantic recovery. For example, as indicated by the red circles in the recovered images, the JTSC scheme correctly reconstructs an arm, while ALL-Selection generates an inaccurate arm due to redundant semantic information in that region.
\section{Conclusions}\label{conclusions}
In this paper, we proposed the JTSC scheme for GAI-assisted semantic image transmissions, where SUs adaptively extract and select cross-modal semantic features from the raw image depending on their NOMA transmission capacities. We maximized the weighted performance of semantic recovery accuracy and transmission latency by jointly optimizing the semantic feature selection strategy, the NOMA decoding order, and the receive beamforming. We designed a cross-modal matching method to quantify the importance of the semantic features, enabling the pruning of low-importance features to reduce the search space. Moreover, we proposed the IM-PPO framework to solve the complex optimization problem involving the black-box component. Depending on the availability of model information, the transmission control variables are tackled by the model-based optimization methods, while the semantic feature selection strategy is learned by the PPO. Numerical results validated that the JTSC scheme achieves superior performance in both semantic accuracy and latency, and the IM-PPO framework significantly enhances learning efficiency compared to benchmark methods.

\footnotesize
\bibliographystyle{IEEEtran}
\bibliography{reference}
\end{document}